\documentclass{article}
\usepackage[colorlinks,bookmarksopen,bookmarksnumbered,citecolor=red,urlcolor=red]{hyperref}
\usepackage{url}
\usepackage{amsmath,amsthm,amsfonts}
\newtheorem{example}{Example}
\newtheorem{remark}{Remark}
\usepackage{xcolor}
\usepackage{multirow}
\usepackage{graphicx}
\usepackage{dirtytalk}
\newcommand\BibTeX{{\rmfamily B\kern-.05em \textsc{i\kern-.025em b}\kern-.08em
T\kern-.1667em\lower.7ex\hbox{E}\kern-.125emX}}
\def\volumeyear{2016}

\newcommand{\new}[1]{\textcolor{black}{#1}}
\newcommand{\ceil}[1]{\left\lceil #1 \right\rceil}

\begin{document}
% \runninghead{Rosenfeld, Shapiro and Talmon}
\title{Proportional Ranking in Primary Elections: A Case Study}
\author{Ariel Rosenfeld$^{1}$, Ehud Shapiro$^2$ and Nimrod Talmon$^3$\\\\
$^1$Bar-Ilan University, Israel\\
$^2$Weizmann Institute of Science, Israel\\
$^3$Ben-Gurion University, Israel
}
\date{Preprint of \url{https://doi.org/10.1177/13540688211066711} (published in \textit{Party Politics}, January 2022)}
% \runninghead{Blind for review}
% \title{Proportional Ranking in Primary Elections: A Case Study}
% \author{Blind for review}
\maketitle

\begin{abstract}
Many democratic political parties hold primary elections, which nicely reflects their democratic nature and promote, among other things, the democratic value of inclusiveness.
However, the methods currently used for holding such primary elections may not be the most suitable, especially if some form of proportional ranking is desired.
In this paper, we compare different algorithmic methods for holding primaries (i.e., different aggregation methods for voters' ballots), by evaluating the degree of proportional ranking that is achieved by each of them using real-world data.
In particular, we compare six different algorithms by analyzing real-world data from a recent primary election conducted by the Israeli \emph{Democratit party}.
Technically, we analyze unique voter data and evaluate the proportionality achieved by means of cluster analysis, aiming at pinpointing the representation that is granted to different voter groups under each of the algorithmic methods considered.
Our finding suggest that, contrary to the most-prominent primaries algorithm used (i.e., \textit{Approval}), other methods such as \textit{Sequential Proportional Approval} or  \textit{Phragmen} can bring about  better proportional ranking and thus may be better suited for primary elections in practice.
\end{abstract}
% \keywords{Primary elections,  proportional ranking, representation, case study}

% \nimrod{let's change "proportional representation" to "proportional ranking (in primaries"}

\section{Introduction}

Parties in many democracies around the world
hold primary elections (e.g., Israel, Canada, UK, Spain, Italy); in such elections -- which we refer to as \emph{primaries} -- party members, and sometimes even sympathizers (without officially being party members), can select and rank their leaders and/or legislative candidates~\cite{astudillo2020sometimes,cross2015politics}. Studies estimate that between $30\%$ to $50\%$ of parties in western democracies use primaries to select their leaders~\cite{cross2014selection} and between $25\%$ to $33\%$ of the more prominent parties in democracies make use of primaries to rank their legislative candidates \cite[page~47]{cross2016promise}.

In particular, ranking a party's legislative candidates is a task often addressed via party primaries, and is the corner stone of the current work. A standard process of party primaries consists of the following two main phases:
  First, eligible voters report their preferences over the candidates, commonly by providing a subset of up to $k$ candidates they \say{approve}; we will refer to these subsets as the voters' \textit{approval ballots}. \new{Given voters' approval ballots, some aggregation method is applied in order to derive a ranked list; we will refer to this aggregation process as a \textit{ranking}. Most prominently, the \textit{Approval} aggregation method is adopted, through which} candidates are ranked by the number of votes they receive.\footnote{Unlikely ties may be broken using a secondary ranking criteria such as alphabetically or at random.} For illustrative purposes, consider Example \ref{exam:app}.
  
\begin{example}\label{exam:app}
Assume a primaries with $3$ available candidates: $a$, $b$, and $c$; and $5$ voters, three of which approve $a$ and $b$ (i.e,, their approval ballots consist of $a$ and $b$) while the other two voters approve $b$ and $c$. Indeed, $a$ gets $3$ approvals, $b$ gets $5$ approvals, while $c$ gets $2$ approvals. Thus, when using Approval as the aggregation method, the result of the primaries would be the ranking \new{$b > a > c$ (i.e., $b$ in the first place, $a$ second, and $c$ last)}.
\end{example}
  
While Approval voting is indeed a natural and simple aggregation method of choice, recently, literature in computational social choice (most particularly the line of work on justified representation~\cite{aziz2017justified}; see the Related Work for more discussion) have demonstrated severe potential shortcomings of this approach as illustrated in Example \ref{exam:problem}.

\begin{figure}[t]
\centering
\includegraphics[width=10cm]{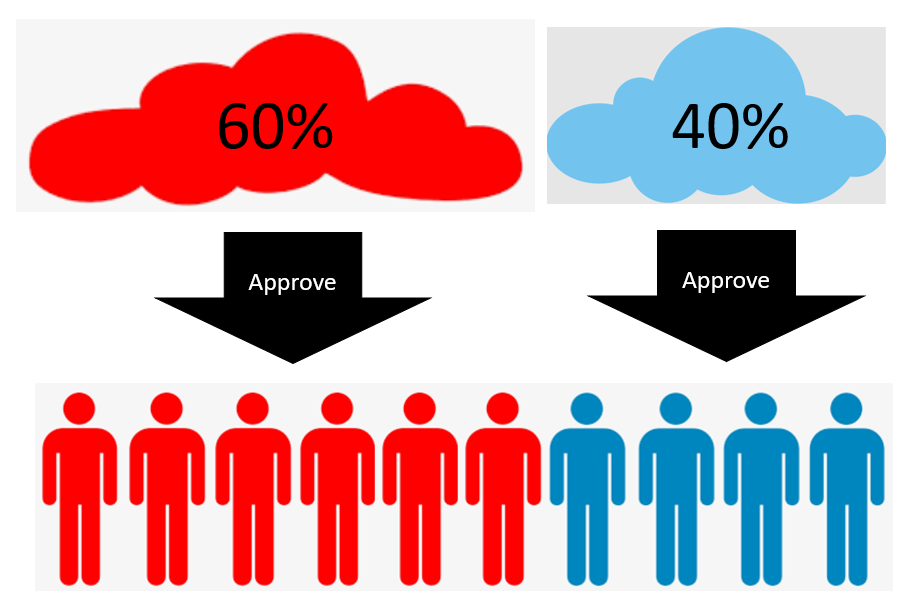}
\caption{Illustration of Example~\ref{exam:problem}.}
\label{figure:figure1}
\end{figure}

\begin{example}\label{exam:problem}
Assume a party with two distinguished groups of voters: say, 60$\%$ of the voters are \emph{red}, while the other 40$\%$ of them are \emph{blue}. Furthermore, assume that also the candidates can be partitioned into red and blue candidates, and that all voters approve the candidates of the same color as they are, namely, all red voters approve all red candidates (and only them), while all blue voters approve all blue candidates (and only them).
Indeed, if Approval voting is adopted, then this would mean that all the red candidates would be ranked before all the blue candidates. Taken together with the number of seats the party would win in the elections, it might eventually mean that only red candidates will represent the party in the parliament while red voters constitute only 60\% of the primaries voter pool. See Figure~\ref{figure:figure1} for an illustration.
\end{example}

Example~\ref{exam:problem} demonstrates an extreme lack of proportionality or representativeness in the outcome of the primaries when adopting Approval voting. \new{Such lack of proportionality may lead the underrepresented groups -- in our case, the blue members -- to feel that the ranked candidate list does not adequately represents them -- in our case, does not represent them at all in parliament. }
Indeed,~\cite{rahat2008democracy} argue that one of the most devastating shortcomings of primaries is the possible existence of a trade-off between the democratic values of ‘inclusion’ and ‘representation’. That is, by increasing the inclusiveness value through primaries, one may make it more difficult for traditionally under-represented groups (e.g., the blue group in Example~\ref{exam:problem}) to be selected and be properly represented as compared to more exclusive selection methods. 

To mitigate this concern, several competing aggregation methods have been recently proposed in the computational social choice literature (in particular~\cite{skowron2016proportional}; see the Related Work for more literature pointers). The core guiding principle of these aggregation methods is that, intuitively, \say{large and cohesive enough voter subgroups should be adequately represented}. Indeed, this intuitive goal is completely violated in the case presented in Example \ref{exam:problem}.
Importantly, however, despite the major theoretical advances in the field of computational social choice regarding rules that aim at achieving such proportionality guarantees, the methods developed by the community have yet to be applied and investigated in a real-world setting or using real-world data. The lack of empirical evidence as to the benefits and limitations of these proposed methods may thus raise doubts regarding their potential use in practice. \new{This study focuses on this gap.}

\new{In particular, }in this work, we provide a unique perspective on the matter in question by using real-world primaries data from the Israeli \emph{Democratit} party\footnote{\url{https://www.democratit.org.il/}} that held such primaries recently. The data, corresponding to approval ballots of $4,507$ voters over $59$ candidates ($37$ male, $22$ female), allows us to investigate the possible advantages and limitations of applying different proportional ranking aggregation methods in the real-world, which is what we set to do in this work. 

Our investigation demonstrates that the use of advanced aggregation methods such as \textit{Proportional Approval voting} or \textit{Phragmen} (which will be explained later in this article) can significantly improve the proportional representativeness of primary election results. Technically, by doing a careful cluster analysis -- in which we partition voters into \say{similar} groups and quantitatively analyze the  representation that each of them gets in the resulting ranking -- we are able to compare different aggregation methods; we conclude that, indeed, these more advanced aggregation methods can significantly promote the notion of proportional ranking.

We view our results as concrete evidence that the use of proportional methods -- such as Sequential Proportional Approval or Phragmen -- in the context of primary elections, can bring about significantly more representative outcomes in real-world primary elections while allowing for inclusive primaries. That is, by adopting these methods, the trade-off between inclusion and representation can be significantly relived and the representation value could be promoted \textit{not} at the expense of the inclusion value. This may be especially important for parties whose ideology is inline with such proportionality values and those in which minority groups are under-represented using the standard Approval method.

% could be better represented and the primaries outcomes would better satisfy the proportionality desideratum.

\subsection{Related Work}

% cross2012politics

The process of intra-party democratization is evident both in leadership selection \cite{pilet2014selection} as well as the selection (and ranking) of legislative candidates \cite{sandri2015party}. These -- as well as other measures -- are implemented in order to make the internal decision-making processes more \say{attractive} and \say{inclusive}   \cite{borz2020contemporary,ignazi2020four}.
% as well as help stop the further erosion of trust citizens attach to political institutions \cite{israel_2021}.
At the same time, however, a trade-off between the democratic values of ‘inclusion’ and ‘representation’ emerges~\cite{rahat2008democracy}. That is, by adopting a primaries system, minority groups may be severely under-represented.  Recently,~\cite{astudillo2021party} have shown that this is in fact the usual case for female political leaders. 
% As a result, the different democratic values of inclusion and representatives are ‘unlikely to be simultaneously maximized in a single institution’.

% The topic of primaries is studied extensively, including several books (see, e.g.,~\cite{boatright2014congressional},~\cite{hirano2019primary},~\cite{sandri2015party}).

Focusing on the issue of representation in primaries,~\cite{kaufmann2003promise} investigates whether holding open primaries increases the representativeness of various factors such as age.
\cite{sides2020representativeness} and~\cite{ranney1968representativeness} study whether voters who actually participate in primary elections are indeed a good sample of the overall party voter population. 
% note that the issue that we are considering in this paper is different, and we refer to it thus as proportionality and not as representativeness.
%
We emphasize that our concern in this paper is regarding the representativeness of different voter groups not with respect to the set of party voters, but their representation in the \emph{outcome} of the primaries. 
% Indeed, in a way, our main claim in this paper is that, in order to improve such proportionality, it is very useful to change the voting rule itself to a more proportional one.

From a computational social choice perspective,~\cite{borodin2019primarily} have demonstrated the potential benefits of holding primaries after suitable mathematical modeling and reasoning techniques as well as computer-based simulations. 
% We are not concerned with such issues, but aim centrally at improving the results of such primaries when issues of proportionality are of importance. 
% Still from the computational social choice perspective, we mention some work on proportionality. 
Focusing on the issue of proportionality and representation, the influential paper of~\cite{aziz2017justified} defines the basis for an axiomatic perspective of proportional representation; in essence, a voting method is said to satisfy proportionality if each sufficiently-large and sufficiently-cohesive group of voters is not completely disregarded by the voting method. This definition draws its inspiration from prior literature on adequate committee elections methods (see~\cite{mwchapter} for an overview).  
\cite{skowron2016proportional} have adapted this notion of proportionality to the setting of proportional \textit{rankings}, in which the result of the election is a ranked list of winners, which is indeed the case in primaries for legislative candidates.
\cite{shapiro2020electing} have expanded on~\cite{skowron2016proportional} and commented on the usability of their methods.

Overall, while studies from both the political science and computational social choice realms have considered the issue of proportional representation, yet, to the best of our knowledge, no study \new{has} yet to provide concrete evidence as to the possible benefits and limitations of proportional ranking methods in real-world party primaries. As such, out work fills an important gap in the literature of both fields.
\new{We do mention the vast amount of literature dealing with the effects of electoral system choice on political representation, see  \cite{karvonen2011personal} for an extensive overview and analysis of the matter.
Given the different cultural, social, and political properties of different countries, some studies have focused on specific countries such as  Russia~\cite{moser2001unexpected} and Brazil~\cite{mainwaring1991politicians}.
% Lijphart~\cite{lijphart1990political} have considered the political effects of the choice of electoral laws.
More related to our work is  \cite{grofman2006impact} who studied the effects of electoral system choice on political parties. In this work, we consider the existing electoral system in Israel as a given.}

Our study focuses on real-world data obtained from an Israeli political party. Extensive literature have been devoted to Israeli Politics including the study of Israeli primaries; see~\cite[Chapter~1]{hazan2021oxford} and the references therein). For example, very recently~\cite{kenig2021candidate} examined the possible relation between candidate primaries expenses and their success in Israeli parties. Notably, the popular systematic analysis of candidate selection methods, including primaries, by~\cite{hazan2010democracy}, also heavily relies on extensive data from Israel. As mentioned before, to the best of our knowledge, ours is the first work to investigate proportional ranking methods using real-world data, both in Israel and in general.

\section{Proportional Ranking}

We first set the jargon and notations needed for our study and define the setting of proportional ranking.
Then, we formally define the aggregation methods we consider in this article.

\subsection{Defining Proportional Ranking}

We formally define the setting of proportional ranking -- which is, at least in the context of the current article, equivalent to the setting of primary elections for legislative candidates; we largely follow the definitions of~\cite{skowron2016proportional}.

Let $C = \{c_1, \ldots, c_m\}$ be a set of $m$ candidates and let $V = \{v_1, \ldots, v_n\}$ be a set of $n$ voters.
Voters provide \emph{approval ballots}, that is, each $v_i$ corresponds to some subset of $C$ (containing those candidates that $A$ approves of); slightly overloading notation, we set $v_i \subseteq C$.
An \emph{election} is thus a tuple $E = (C, V)$.

An \emph{aggregation method} takes as input an election $E = (C, V)$ and outputs a ranking over $C$. For an aggregation method $\mathcal{R}$, we denote the output of $\mathcal{R}$ on an input election $E$ by $r := \mathcal{R}(E)$, so $r$ is a linear order of~$C$. We write $r$ as a vector $r = (r_1, \ldots, r_m)$ such that $r_i \in C$ is the candidate ranked at position $i$ (e.g., $r_1$ is the candidate at the first position by applying $\mathcal{R}$ on the election $E$).

\subsection{Methods for Proportional Ranking}

We focus on the following six aggregation methods. For each of the methods we ignore the unlikely case of ties (which was not observed in our data as later discussed. Indeed, ties may require special attention such as a secondary ranking criteria).
Note that we have three \say{basic} aggregation methods, and for each of those we have two variants: one without diversity constraints, and another with diversity constraints. In particular, those three diversity-enforcing aggregation methods make sure that there is no prefix larger than 1 of the outcome ranking in which there is a strict male majority. Formally, they will make sure that for every $k>1$ there will be more than $\ceil{\frac{k}{2}}$ male candidates in the first $k$ positions of the ranking.  

\begin{description}

\item 
\textbf{Approval}:
  Approval is probably the most widely used method.
For a candidate $c \in C$, and given some $V$, we write $S(c) := |\{v \in V: c \in v\}|$ to be the number of voters approving $c$; we refer to this number as the \emph{approval score} of $c$.
The Approval method orders the candidates by decreasing order of their approval scores.

\item
\textbf{DApproval}:
  DApproval is similar to Approval but with enforced gender diversity or quotas -- in particular, it guarantees the following diversity constraint: ``there shall not be a male strict majority in any prefix larger than 1 of the resulting ranking''. We adopted this gender diversity constraint as this was decided by the constitution of the Israeli Democratit party. Naturally, other diversity constraints may be implemented as well.
In practice, DApproval operates in iterations. Initially, we have $r = ()$; then, in each iteration, we append to $r$ the candidate with the highest approval score remaining, as long as we do not violate the diversity constraint.

\item
\textbf{SPAV}:
  SPAV stands for Sequential Proportional Approval Voting.
The operation of SPAV can be explained as follows: Initially, the \emph{weight} of each ballot is $1$; then, in each iteration we select a new \say{winner} to be ranked in the next position. In each iteration we select the candidate for which the weighted sum of the ballots approving it is the highest.
We refer the interested reader to the paper of~\cite{skowron2016proportional} or to the paper of~\cite{shapiro2020electing} for more formal as well as intuitive descriptions of SPAV.

\item
\textbf{DSPAV}:
  DSPAV is similar to SPAV but with enforced gender diversity as discussed before. 
Technically, DSPAV operates similarly to SPAV, but, in each iteration we make sure that the diversity constraint is indeed satisfied. Practically, in each iteration we select the candidate with the highest weighted sum of votes as long as the constraint is respected.  

\item
\textbf{Phragmen}:
\new{Phragmen is an adaptation of the voting rules proposed by the Swedish mathematician Lars Edvard Phragmen over 120 years ago \cite{phragmen1,phragmen2,phragmen3,phragmen4} for electing unranked committees. These rules were recently popularized by~\cite{brill2017phragmen}.}
Phragmen is a somewhat complex aggregation method compared to the previous ones.
\new{The main idea behind the method of Phragmen is to identify a ranked committee such that the support that it gets from the voters is distributed as evenly as possible among the electorate.}
Perhaps the most intuitive explanation for the operation of this method is \new{via the metaphore of ``bank accounts'':
  Imagine that each voter has her own ``bank account''.
  Now, assume that there is a continuous flow of money being deposited to all accounts such that, as time advances, the bank account of each voter is accumulating slightly more money (in particular, a continuous infinitesimal increase of money). Given such continuous flow of money, Phragmen works by adding candidates to the ranked list, starting initially from an empty ranked list (i.e., $r = ()$);
  then, whenever there is a candidate for whom the sum of money accumulated by voters that approve that candidate is equal to $1$, then this candidate is appended to the result and the bank accounts of those voters who approved that candidate are emptied. The algorithm halts after all candidates have been appended and ranked, and outputs the corresponding final ranking.}
\new{We refer the interested reader to technical report of \cite{janson2016phragmen} for further details.}

\item
\textbf{DPhragmen}:
  DPhragmen is similar to Phragmen but with enforced gender diversity.
Technically, DPhragmen operates similarly to Phragmen, but, in each iteration we make sure that the diversity constraint is indeed satisfied.

\end{description}

\section{Data}\label{sec:data}

In order to examine the possible advantages and limitations of adopting a proportional ranking approach, we obtained data from the Israeli Democratit party. 
Next, we first discuss the Democratit party and its primaries process as was held in 2020. Then, we analyze the data, focusing on identifying similarities between voters and clustering them into cohesive groups. These groups will be used for the comprehensive comparison of the examined aggregation methods.

\subsection{The Democratit Party}

The Democratit party is an Israeli political party. The slogan of the party is the Hebrew translation of the ``slogan'' of the French revolution, i.e., Liberte, Egalite, Fratarnite.

One of the underlying principles of the Democratit party is transparency of decision making and a strive towards implementing direct democracy constructs, perhaps towards a truly direct democracy internal decision making process.
The party is rather young (as it was founded in 2020) and its specific goals are continuously being debated and changed; this is mainly so as all party members have a say in the party decisions. 

The data we analyze in the current article is from the first primary elections held in the Democratit party; these elections effectively chosen the party leadership as well. At the end, the party chose not to run for the general 2021 Israeli legislative election but considers doing so in the next elections.

\subsection{The Primaries in the Democratit Party}

As obliged by the constitution of the party, the 2020 primaries in the Democratit party were held as follows:

\paragraph{Elicitation Method.}
In order to decide on a candidate list for the Israeli general elections, the Democratit party decided to use a ($\leq10$)-Approval ballots, where each voter selected up to $10$ candidates from the list of candidates (where each party member had the possibility of running for the primaries).

\paragraph{Candidates and Voters}
The primaries consisted of $59$ candidates -- $37$ male and $22$ female candidates -- and $4507$ voters. No other pieces of information were provided to us.

\paragraph{Aggregation Method.}
The party decided to use Diversity-aware-SPAV, DSPAV in short, and announced its decision in its bylaws before the primaries took place.

\section{Basic Data Analysis}

First, we provide some basic analysis.
In particular, we start by reporting on the distribution of the ballot sizes, and the examination of ballots \say{uniqueness}.

\paragraph{Ballot sizes}

\begin{figure}
\centering
\includegraphics[width=0.88\linewidth]{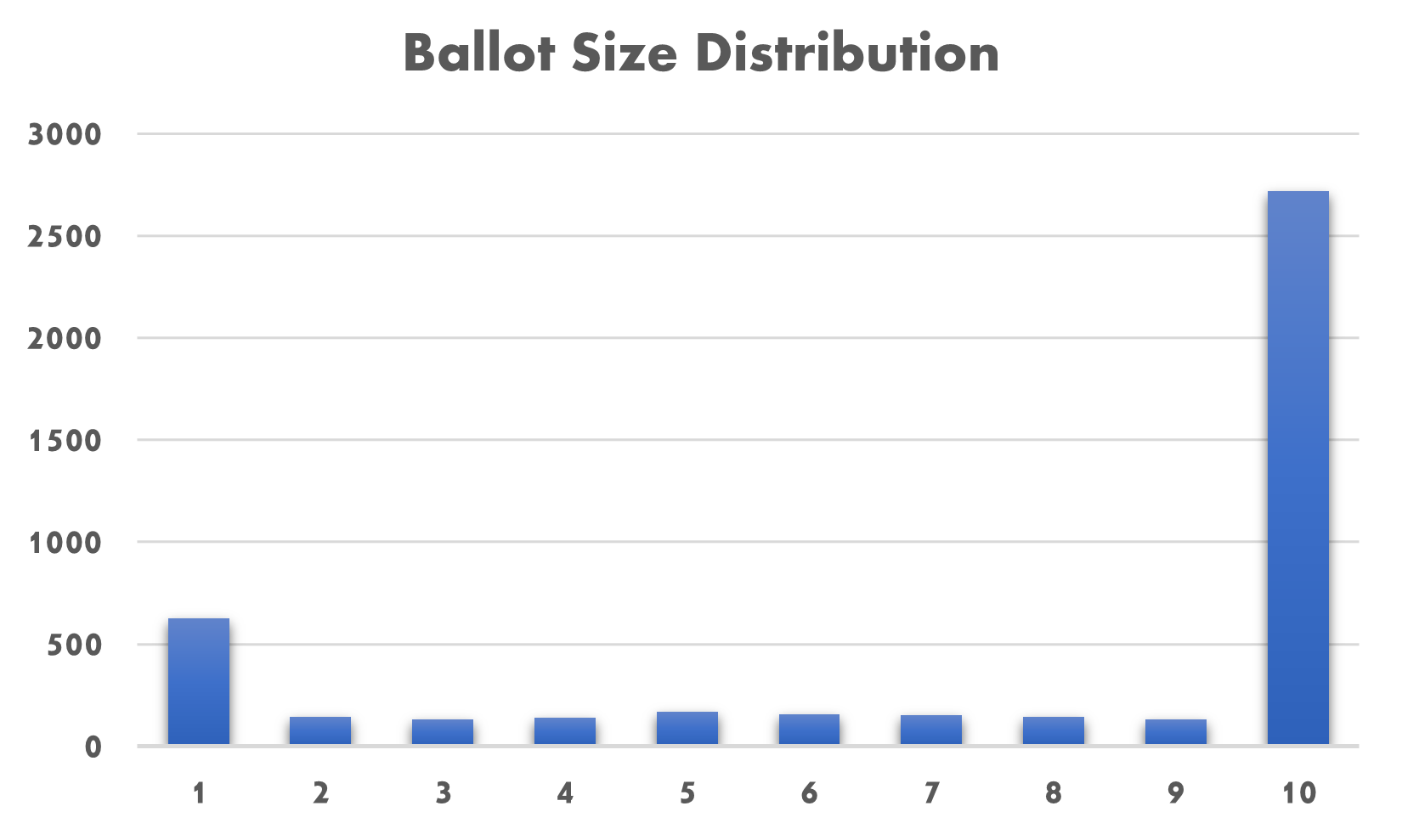}
\caption{Ballot size distribution. The X-axis marks the size of the ballot and the Y-axis marks its frequency.}
\label{fig:hist}
\end{figure}

The distribution of ballot size is presented in Figure \ref{fig:hist}. Slightly more than 60\% of the ballots were \say{full} ballots, approving exactly 10 candidates. Surprisingly (at least to us), 13.8\% of the ballots approved only a single candidate. The remaining 26.2\% of voters approved between 2 and 9 candidates with the size of the ballots distributed almost uniformly across this range. 
  
\paragraph{Ballot Uniqueness}

Out of $4,507$ approval ballots, $3,713$ (82.3\%) where \textit{unique}. That is, the vast majority of voters' approval ballots do not fully coincide with each other. 
Analysing the non-unique ballots that have appeared at least $5$ times in the ballot set (i.e., approximately 0.1\% prevalence) shows that there are only $38$ such ballots with the majority of which being very small. 
Specifically, $31$ ballots were of size $1$ (i.e., voting for only a single candidate) and appeared more than $600$ times, $3$ ballots were of size $2$ (i.e., voting for only two candidates) and appeared more than $100$ times, $2$ ballots of size $5$ and $2$ ballot profiles of size $10$ which together appeared less than $100$ times. The average size of these non-unique ballots is $1.76$. Interestingly, the top $3$ most prevalent ballots were of size $1$.

\begin{remark}
The ballot size and similarity analysis leads us to the realization that most voters are ``independent'' and vote on their own merits rather than cast the same ballot many people agreed upon.
\end{remark}

\section{Comparison between Possible Results}

First, let us look at the primaries outcomes that would have results  from the application of each of the aggregation methods we consider.
Indeed, Table~\ref{table:orderings} shows the top ranked $20$ people under each of the aggregation methods we consider.

\begin{table*}[t]
\centering
\begin{tabular}{ c|c|c||c|c|c }
\hline
Approval & SPAV & Phragmen & DApproval & DSPAV & DPhragmen \\
 \hline
c1&c1&c1&c1&\bf{c1}&c1\\
c2&c2&c2&c2&\bf{c2}&c2\\
c3&c3&c3&c3&\bf{c3}&c3\\
c6&c4&c4&c6&\bf{c4}&c4\\
c5&c5&c6&c5&\bf{c5}&c5\\
c4&c6&c5&c4&\bf{c6}&c6\\
c8&c8&c8&c11&\bf{c7}&c7\\
c10&c10&c10&c8&\bf{c8}&c8\\
c11&c12&c7&c7&\bf{c9}&c11\\
c16&c7&c12&c10&\bf{c10}&c10\\
c7&c9&c11&c9&\bf{c11}&c9\\
c9&c11&c9&c16&\bf{c12}&c12\\
c14&c14&c16&c13&\bf{c13}&c13\\
c20&c18&c14&c14&\bf{c14}&c16\\
c12&c16&c20&c19&\bf{c15}&c17\\
c22&c20&c18&c20&\bf{c16}&c14\\
c24&c22&c22&c17&\bf{c17}&c19\\
c28&c24&c28&c12&\bf{c18}&c20\\
c18&c26&c24&c15&\bf{c19}&c15\\
c13&c13&c13&c22&\bf{c20}&c28\\
\hline
\end{tabular}
\caption{The primaries results under the six examined aggregation rules. On the left, the three methods which do not force diversity. Recall, DSPAV was actually implemented by the party, hence the results are highlighted for reference.}
\label{table:orderings}
\end{table*}

\subsection{Differences and Interpretation}

Let us look at the results of Table~\ref{table:orderings}.
Indeed, all methods coincide on the first three places, possibly indicating that these three candidates are in the general consensus and also consist of both male and female candidates.

Differences begin on the 4th place. We mention that, in Israel, $4$ seats are the minimal threshold for entering the parliament, so the 4th place is very relevant. In words, when transitioning from (D)Approval as a baseline, to either (D)SPAV or (D)Phragmen, candidate $c_6$ drops either one or two spots in the resulting ranking. This indicates that the voters who supported $c_6$ are very likely to support candidates from the set $\{c_1,c_2,c_3\}$; as such, these voters are already \say{represented} by the first three candidates, while voters approving $c_4$ and $c_5$ were significantly less represented by  $c_1,c_2$ and $c_3$.
Indeed, the data supports this conclusion. Starting with $c_6$, she was approved by 1,522 voters, of whom, 1,448 (95\%) also approved one of the candidates $c_1,c_2,c_3$. $c_4$ and $c_5$, on the other hand, were approved by less voters -- 1,475 and 1,500 voters, respectively, yet of these voters, approximately 89\% have approved one of the candidates $c_1,c_2,c_3$. It is important to note that candidates $c_1,c_2,c_3$ are very popular in the primaries, with 80\% of all voters approving at least one of them. 
% this can be inferred both from the definition of ``large and cohesive'' groups of voters and also from the algorithmic description of SPAV and Phragmen.

% \nimrod{Regarding the above paragraph: It is somewhat a very important paragraph, but it is slightly ``hidden''; in particular, maybe it is hard for the reader to understand the relation to the abstractly-defined notion of ``large and cohesive'' groups? One remedy may be to speak directly of the definition of JR -- but this might be to hard for our readers perhaps.. so maybe to say who are the groups? I'm not sure..}
% \ariel{compute: number of voters approving 1,2,3 that also 4 against those approving 1,2,3 that also 6}

Interestingly, when diversity is not enforced, SPAV and Pharagmen differ on the 5th and 6th places while in the diversity-enforced case, they differ only from the 9th position. 
\new{When diversity is enforced, indeed, there is no strict male majority in any prefix smaller than $44$. Note that since there are only $22$ female candidates, one cannot enforce diversity after the $44$th place.}

% SPAV and Phramen are very similar and so are the DSPAV and DPharmen (yet, they do not coincide). This is aligned with the rational behind these approaches, as both seek to bring about better representation for minority groups. 

\subsection{Quantification of the Differences}

In order to quantify the differences between the six possible rankings of Table \ref{table:orderings}, we use the popular Kendall $\tau$  distance. $\tau$ is a normalized metric that relies on the number of pairwise disagreements between two ranking lists. Namely, it calculates the number of times a pair of candidates (say, $c_i$ and $c_j$) are not ranked in the same way in both rankings (i.e., $c_i$ is ranked higher than $c_j$ in the first ranking but not in the second ranking, or vise versa). The $\tau$ rank distance is normalized (i.e., it falls between 0 and 1), with 0 reading \say{perfect agreement} and 1 reading \say{perfect disagreement}. In particular, the larger $\tau$ is, the more dissimilar the two examined rankings are. As can be seen from Table \ref{table:tau}, Approval and Phragmen seem to be the most similar pair of rankings. Surprisingly, even when diversity is enforced on Phragmen (i.e., DPhragmen), Approval and DPhrgmen are the second most similar pair of rankings. On the other hand, when considering SPAV, it significantly differs from Approval, with and without the diversity constraint. These results are somewhat surprising as both SPAV and Phragmen strive to promote proportional at the expense of plurality. 

\begin{table*}[t]
\centering
\begin{tabular}{ c||c|c|c|c|c }
\hline
Method & SPAV & Phragmen & DApproval & DSPAV & DPhragmen \\
\hline 
Approval & 0.5 &\textbf{0.073} &0.347 &0.373 &0.247\\
\hline
SPAV & - & 0.468 & 0.531 & 0.378 &0.473   \\
 \hline
Phragmen & - & - & 0.347 & 0.384 & 0.473 \\ 
 \hline
DApproval & - & -& -& 0.3 & 0.321 \\
 \hline
DSPAV &  - & -& -& -& 0.431\\
\hline
\end{tabular}
\caption{Kendall $\tau$ distance between the six examined aggregation methods. In bold, the most similar pair of rankings.}
\label{table:tau}
\end{table*}

\section{Clustering-based Comparison}

Next we perform a comparison between the aggregation methods based on a cluster analysis.
\new{Cluster analysis is the mathematical task of grouping (i.e., clustering) a set of objects such that objects in the same group would be as similar as possible while objects from different groups would be as dissimilar as possible. It is often used to discover the underlying structure in a set of objects; correspondingly, in our setting, we use cluster analysis in order to identify groups of similar and dissimilar voters.}
First we describe the general idea and the specific clustering techniques used, and then we present and discuss the results.

\subsection{Clustering Techniques}

Recall that the basic premise of the methods that promise some proportional ranking is to promote the adequate representation of \say{large and cohesive} groups of voters. Since the vast majority of voters do not vote exactly the same way (specifically, recall that $82.3\%$ of the ballots were unique), we shall define a measure of similarity between voters -- to be able to consider groups of voters, thus below we discuss several distance metrics. Then, using the introduced metrics, we are indeed able to cluster the voters based on their similarity.
Later on we use these clusters also for comparing the results of the different aggregation methods.

\subsection{Voter Similarity}

In the context of this work, voters are represented by their approval ballots. As such, similarity between two voters shall be defined as similarity between two approval. While there are many possibilities to define such metric, here we focus on two standard metrics commonly used when comparing groups in their mathematical meaning:

\begin{enumerate}

\item \emph{Hamming Distance} (denoted by $H$):
  $H$ measures the minimum number of substitutions required to change one ballot into the other. Namely, the minimum number of times one needs to approve a new candidate or disapprove an existing candidate in order to transform one ballot to another. In particular, given two ballots $v_i,v_j$, $H(v_i,v_j)=0$ if and only if $v_i=v_j$. Otherwise, the higher $H(v_i,v_j)$, the more dissimilar voters $i$ and $j$ are.
  For example, if voter $v_1$ approves $a$ and $b$ while voter $v_2$ approves $a$ and $c$ then $H(v_1, v_2) = 2$.
  
\item \emph{Jaccard Index} (denoted by $J$):
  $J$ measures the similarity between two ballots $v_i$ and $v_j$ by dividing the size of the intersection between the two by the size of their union. For example, if $v_i=\{c_j,c_k\}$ and $v_j=\{c_j,c_l\}$ then $J(v_i,v_j)=1/3$ since only $c_j$ is in their intersection while $c_j,c_k,c_l$ are in their union. In particular, $J(v_i,v_j)=1$ if and only if $v_i=v_j$. Otherwise, the lower the value of $J(v_i,v_j)$ is, the more dissimilar the two voters are and $J(v_i,v_j)=0$ if and only if $v_i$ and $v_j$ are disjoint. 
  
\end{enumerate}

\subsection{Clustering}

Using the two metrics described above, we are able to cluster the voters into groups. To this end, we use the standard K-Means algorithm~\cite{macqueen1967some} with $k$ ranging from $1$ to $10$.
In simple words, the K-Means algorithm -- in our context -- aims at partitioning the ballots into $k$ clusters in which each ballot belongs to the cluster with the nearest mean, which, in turn, serves as a prototype or representative of the cluster.

In order to determine the appropriate number of clusters ($k$) we use the classic Eblow Method (which can be traced back to the mid-20th-century~\cite{thorndike1953belongs}).
Simply put, the Elbow method is a heuristic used in determining the number of clusters that underline a data set by plotting the explained variation in the data as a function of the number of clusters, and picking the \say{elbow} of the curve as the number of clusters to use. 

\begin{figure}
\centering
\includegraphics[width=0.75\linewidth]{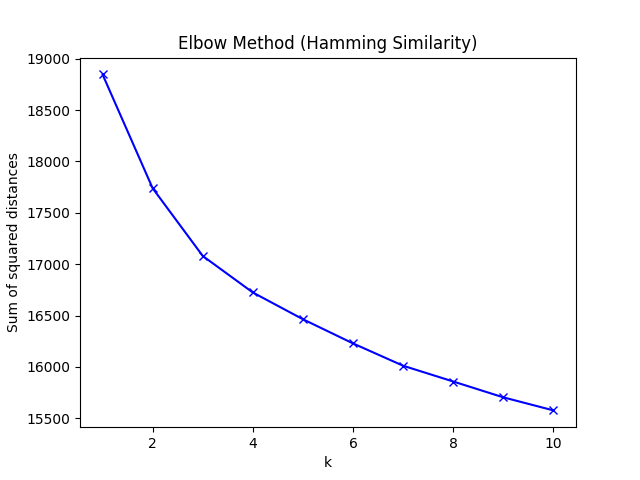}
\includegraphics[width=0.75\linewidth]{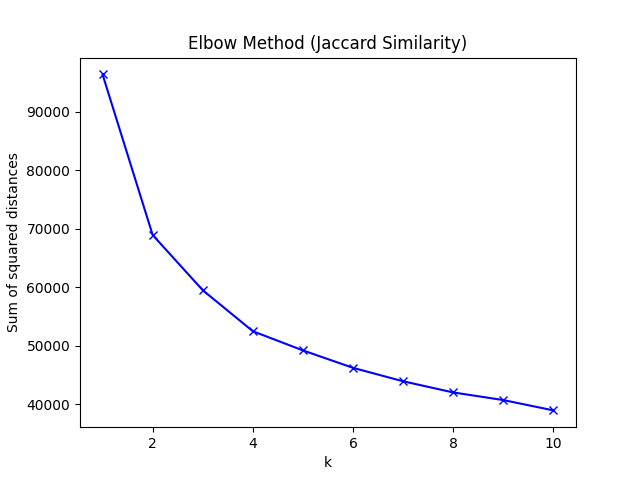}
\caption{Elbow method graphs using Hamming distance (top) and Jaccard index (bottom). }
\label{fig:elbows}
\end{figure}

As can be seen in Figure~\ref{fig:elbows}, the application of the Elbow Method under the two examined metrics suggests the use of $k = 3$ in both cases. It is important to note that the Elbow Method is inherently heuristic and so the exact position of the \say{elbow} may be subject to interpretation and debate. The analysis presented next, however, is only mildly effected by selecting other reasonable values for $k$ such as $k = 4$ under the Jaccard Index metric. 

Recall that we have divided our voters into three groups earlier in this article. Using these groups, we now turn to investigate how the different methods effected these groups in terms of their representation. To this end, we first define three measures to capture each group's \say{satisfaction} from its representation in a given ranking. Given a group of voters~$V'$ and a ranking $r$ we consider the following functions;

\paragraph{TOP-4}
Since the Democratit party was expected to struggle to pass the threshold for entering the parliament (recall that the current threshold in Israel is $4$), one interpretation for \say{satisfaction} would be to measure the average number of approved candidates each voter group gets in the $4$ top-ranked candidates. Formally, for each voter $v\in V'$, we calculate the number of approved candidates in the 4 top-ranked candidates in~$r$, we will refer to this measure as the \say{voter satisfaction}. Then, for each voter group $V'$, we average  its members voter satisfaction measures to reach a  \say{group satisfaction} score.

\begin{example}
To illustrate the definition of TOP-4 satisfaction consider a set of 13 candidates $\{1, 2, 3, 4, 5, 6, 7, 8, 9, 10, 11, 12, 13\}$ and a subgroup of voters $V'$ consisting of $2$ voters: the first approves $\{3, 4, 13\}$ and the second approves $\{1, 6, 11\}$. Given that the primaries result is the order $\{1, 2, 3, 4, 5, 6, 7, 8, 9, 10, 13, 12, 11\}$, we have that the TOP-4 satisfaction level of $V'$ is $1.5$; the first voter approves two candidates in the top four ranked candidates ($3$ and $4$) and the second voter approves only one ($1$). 
\end{example}

\paragraph{Linear-Top-10}
The \textit{TOP-4} interpretation may be criticized for several reasons. First, it considers only the top $4$ candidates while the output candidate list is much larger. Second, it places no importance on the ranking itself but rather only membership in the top~4. To cope with these limitations, we propose to amend the calculation by first considering the top 10 ranked candidates and not just the top 4 and, in addition, applying a scoring rule such that the higher an approved candidate is, the higher the satisfaction she brings to her supporters. We adopt the following simple scoring rule: the voter satisfaction from approving a top-10 ranked candidate is inversely related to its location in the ranking: the highest ranked candidate brings about 10 \say{points} to her supporters while the $10$th will bring about only a single point. Using this function we consider both the presence of approved candidates in the top 10 places as well as their actual position within the ranking.

\begin{example}
To illustrate the definition of Linear-Top-10 satisfaction, consider a set of candidates $\{1, 2, 3, 4, 5, 6, 7, 8, 9, 10, 11, 12, 13\}$ and a subgroup of voters $V'$ consisting of $2$ voters: the first approves $\{3, 4, 13\}$ and the second approves $\{1, 6, 11\}$. Given that the primaries result is the order $\{1, 2, 3, 4, 5, 6, 7, 8, 9, 10, 13, 12, 11\}$, we have that the Linear-Top-10 satisfaction of the first voter is $8 + 7 = 15$ where the first element is due to the position of candidate 3 and the second due to the position of candidate 4. The satisfaction of the second voter is $10+5=15$ where the first element is due to the position of candidate 1 and the second due to the position of candidate 6. This results in a group satisfaction of 15. 
\end{example}

\paragraph{Linear}
The last function we consider relaxes the previous functions even further by considering a linear factoring over the entire ranked list and not just the top 10 spots. Namely, the highest ranked candidate is factored by $59$ (the number of candidates in the primaries) and the lowest ranked candidate is factored by $1$.

\begin{example}
To illustrate the definition of Liner satisfaction, consider a set of candidates $\{1, 2, 3, 4, 5, 6, 7, 8, 9, 10, 11, 12, 13\}$ and a subgroup of voters $V'$ consisting of $2$ voters: the first approves $\{3, 4, 13\}$ and the second approves $\{1, 6, 11\}$. Given that the primaries result is the order $\{1, 2, 3, 4, 5, 6, 7, 8, 9, 10, 13, 12, 11\}$, we have that the Linear satisfaction of the first voter is $11 + 10 + 3 = 24$ where the first element is due to the position of candidate $3$, the second is due to the position of candidate $4$ and the last is due to the position of candidate $11$. The satisfaction of the second voter is $13 + 8 + 1 = 22$ where the first element is due to the position of candidate $1$, the second is due to the position of candidate $6$ and the last is due to the position of candidate $13$. The resulting group satisfaction level is 23.
\end{example}

\subsection{Satisfaction Levels}

Next, in Tables~\ref{tab:a} and~\ref{tab:b}, we provide the  satisfaction levels for each of the three functions described above, for each of the three identified groups, under each of the different aggregation methods.

\begin{table*}[t]
\begin{center}
\begin{tabular}[H]{|c||c|c|c|c| } 
\hline
Metric & Cluster Size & Approval & SPAV & Phragmen  \\
\hline
% \hline
\multirow{3}{-10pt}{H} 
& 773 & 2.6/26.8/409.9 & 2.5/25.6/404.6 & 2.5/25.9/408.5\\ 
& 991 & 2.2/24.0/412.4 & 2.0/23.1/406.2 & 2.0/23.3/409.2\\ 
& 955 & 1.3/13.9/360.7 & 1.7/15.3/361.9 & 1.7/15.1/361.5\\
\hline
\multirow{3}{0pt}{J} 
& 566 & 2.2/23.0/401.4 & 2.1/21.2/393.0 & 2.1/21.3/393.7\\
& 1108 & 2.0/21.0/391.0 & 2.1/21.1/392.1 & 2.1/21.1/391.8\\
& 1045 & 2.0/21.2/391.6 & 2.0/21.2/393.2 & 2.0/21.2/395.2\\
\hline
\hline
\end{tabular}
\end{center}
\caption{Clustering-based analysis of the results provided by the Approval, SPAV and Phragmen aggregation methods. The top half of the table reports the results under the Hamming metric clustering while the lower half reports the results under the Jaccard metric. Each cell reports three satisfaction levels in the following order: Top-4/Linear-Top-10/Linear.}
\label{tab:a}
\end{table*}

\begin{table*}[t]
\begin{center}
\begin{tabular}[H]{|c||c|c|c|c| } 
\hline
Metric & Cluster Size & DApproval & DSPAV & DPhragmen  \\
\hline
% \hline
\multirow{3}{-10pt}{H} 
& 773 & 2.6/27.6/405.9 & 2.5/25.3/400.1 & 2.5/25.4/402.6\\ 
& 991 & 2.2/22.3/404.2 & 2.0/22.0/398.1 & 2.0/21.8/401.1\\ 
& 955 & 1.3/14.4/353.8 & 1.7/16.3/355.5 & 1.7/16.7/354.8\\
\hline
\multirow{3}{0pt}{J} 
& 566 & 2.2/23.1/398.2 & 2.1/22.6/394.9 & 2.1/22.8/396.2\\
& 1108 & 2.0/21.0/386.4 & 2.1/21.0/387.2 & 2.1/20.9/385.1\\
& 1045 & 2.0/20.7/391.6 & 2.0/21.0/393.0 & 2.0/21.0/395.2\\
\hline
\hline
\end{tabular}
\end{center}
\caption{Clustering-based analysis of the results provided by the DApproval, DSPAV and DPhragmen aggregation methods. The top half of the table reports the results under the Hamming metric clustering while the lower half reports the results under the Jaccard metric. Each cell reports three satisfaction levels in the following order: Top-4/Linear-Top-10/Linear.}
\label{tab:b}
\end{table*}

We analyze the important information of Tables~\ref{tab:a} and~\ref{tab:b} by considering each satisfaction function separately:

\paragraph{Top-4}
As discussed before, the top ranking 4 candidates in SPAV, Phragmen, DSPAV and DPhragmen coincide. As such, it is easy to verify that the clusters' associated top-4 satisfaction across these four aggregation methods is unaltered. Similarly, since the top ranking 4 candidates in Approval, and DApproval coincide, the same argument holds for the associated satisfaction under these two method.
An examination of the satisfaction \say{distribution} shows that the results of (D)Approval display greater levels of inequality between the clusters compared to the results of (D)SPAV and (D)Phragmen. For example, considering the Hamming metric, under (D)Approval, there is a significant inequality with one cluster being \say{extremely unsatisfied} with only 1.3 compared to 2.6 and 2.2. Under the (D)SPAV and (D)Phragmen aggregation methods, this cluster's satisfaction increases to 1.7 while the two other clusters are slightly impaired to allow this increase. The resulting satisfaction levels are obviously distributed more evenly in these rankings.

\paragraph{Linear-Top-10}
Similarly to the TOP-4 analysis, under the Approval and DApproval methods, we see high degrees of inequality between the clusters. When we compare these results to those achieved by (D)SPAV and (D)Phargmen it is evident that both bring about more even distributions of satisfaction across the clusters. However, it seems that (D)SPAV is slightly more aggressive in doing so compared to (D)Phargmen. For example, considering the Hamming metric, the transition from Approval to SPAV will lead the most satisfied cluster (the top one) to \say{lose} 1.2 of its satisfaction level compared to a loss of 0.9 under Phragmen. This excess loss allows SPAV to compensate the under-represented cluster (bottom one) by 1.4 points while Phragmen was able to provide only 1.2 points. This phenomena is consistent across Tables 3 and 4. 
Interestingly, the introduction of the diversity constraint have consistently improved the satisfaction levels of the bottom group under the Hamming metric across all examined aggregation methods. Specifically, this group seems to have approved female candidates at significantly higher level than the other two groups. On the other hand, the opposite effect seems to occur for the second group in the top half of both tables. This means that this group is more associated with approvals to male candidates.

\paragraph{Linear.}
The analysis of the Linear satisfaction function seems to align well with that of the Linear-Top-10 function presented above. Here, too, SPAV and DSPAV seem to provide more equal distributions of satisfaction levels across the groups compared Phargmen and DPhrgmen which, in turn, also promote equality compared to Approval and DAprroval. As was the case before, similar effects emerge when the diversity constraint is enforced. 

% However, it is important to note that the difference in satisfaction levels between the methods seemed to have shrunk significantly in relative terms. Specifically, while the differences under the TOP-4 and Linear-Top-10 functions where relatively large (in one case, the satisfaction of one group was twice as large as another), this is not the case here. Here, the differences seem to be mild in relative terms. The reason behind this result is simple, the voter groups are not homogenous and therefore, almost all candidates recive some level of support within each group.  \textit{all} positions are considered in the calculation of the group's satsfaction. Namely, if a candidate which enjoys high support levels falls below  
% \subsubsection{Bottom Line}

% \ariel{So the bottom line is that indeed SPAV and Phragmen decrease the differences in the satisfaction of the voter clusters ;]}

\section{Discussion and Outlook}

Primary elections for a party's legislative candidates is often preformed in two steps: First, eligible voters provide a subset of candidates they approve of, and then, some aggregation method - most prominently Approval - is applied to derive a ranking. Focusing on the latter phase, in this work, we have used unique voter data from the primary elections held by the Israeli Democratit party and evaluated six aggregation methods, including the extremely popular Approval method, as well as other aggregation methods specifically geared towards proportional representation, using real-world data.

Our results demonstrate, for the first time using real-world data, that alternative aggregation methods such as (D)SPAV and (D)Phragmen can bring about better proportional representation for under-represented groups (as captured by the (D)Approval method).  Specifically, using cluster analysis, we show that the application of (D)Approval voting may lead to high degrees of inequality between voter groups with some of which being severely under-represented. When alternative methods that promote proportional representation are applied, significantly better rankings are achieved in this respect. The above is true both when a diversity constraint is enforced on each method, and without such constraint. These results are summarized in Tables \ref{tab:a} and \ref{tab:b}. \new{We also shared our clustering results with several party officials in order to  better understand the characterises of each cluster. For one of the clusters it was very clear that its members supported Israeli Arab candidates almost exclusively. Israeli Arabs are a minority group in Israel, forming about 21\% of the Israeli population \cite{IL21}. }

It is important to note that the differences between all six method are not dramatic as shown in Table \ref{table:orderings}. Namely, the differences start only at the 4th position in the rankings, indicating that the top 3 ranked candidates are very much in consensus. However, we argue that the encountered differences bare significant importance, especially since the minimal threshold for a party to enter the Israeli Parliament is 4 seats. Differences in other, lower ranking, positions can also have significant impact on candidates' involvement in the campaign and voters' perceptions of the candidates list and their support. (And, of course, other data may reveal different, perhaps more significant, results.)  

Some non-trivial differences between (D)SPAV and (D)Phragmen are encountered in our results: First, it seems that (D)Phragmen results in a ranking more similar to that derived by Approval -- this is especially visible through the use of the Kendall $\tau$ distance (Table \ref{table:tau}). Second, (D)SPAV seems to bring about more balanced groups' satisfaction levels, as shown in Tables \ref{tab:a} and \ref{tab:b}. These two results seem to align, namely, (D)SPAV seems to differ more from Approval thus better achieving proportional representation than (D)Phragmen.

It is important to note that, when comparing two rankings, mixed effects are encountered. For example, when a diversity constraint is enforced, female candidates tend to move up the ranking, but that can only happen at the expense of (probably) more popular male candidates who have to move down the ranking. This may lead to voter groups who are associated with higher approval ratings for female candidates to be satisfied with the former, but unsatisfied with the latter since they still approve, to a certain extent, male candidates as well. Naturally, this trade-off cannot be avoided as only a single candidate may occupy any single position. In the more general context, we view this phenomena as a reminiscent of the inherent conflict between maximizing social welfare (in our case, picking the most popular candidates) and providing fair treatment to minorities (e.g., female candidates). See \cite{peters2020proportionality} for a discussion on the topic.

While we have demonstrated a few key benefits for (D)SPAV and (D)Phragmen over (D)Approval, it is important to note that these, too, have their limitations. \new{First, it is important to note that the examined methods are relevant only when ranking several candidates is necessary (in particular, all of them coincide for settings such as voting for single-member constituencies).} Yet, we believe that the most prominent limitation would be lack of voter understanding of the aggregation method used. Specifically, while Approval has a long tradition and is very natural to use and comprehend, other methods are significantly less popular in practice and less trivial to understand. Adding to this, the resulting ranking is hard to verify without direct access to the approval ballots and an appropriate computer code. \new{Technically, one can \say{run} the methods with a pen and paper, but that might be a rather tedious exercise in real-world settings.} Taken jointly, distrust in the primaries process may arise. The \new{Democratit} party has decided to use DSPAV \textit{prior} to the primary elections. From our communications with party officials, it seems that there was very little resistance to this decision \new{with most members expressing general understanding of the method used. Nevertheless,} in order to maintain transparency, our code was made available to all members. We speculate that more resistance may be encountered in other, more traditional, parties which have used Approval in the past. \new{In addition, the fact that voters knew, in advance, that the DSPAV method will be implemented may have led some of them to try and manipulate their ballots. However, given the inherent complexity of successfully manipulating the results in this real-world settings and the existing literature demonstrating that manipulations tend to be counter-productive for those who practice it (e.g., \cite{roth1999redesign,rosenfeld2020too}), one can reasonably expect most voters to vote the same under the examined methods.}
\new{In addition, even though the choice of DSPAV was publicly announced and was available on the Democratit website, based on a post-factum meeting held with party members and candidates -- the common voter did not pay too much attention to this fact.}

We recognize that the current study is limited by the amount and quality of the data used. In the context of this work, we have used a unique set of $4,507$ voter approval ballots over $59$ candidates, yet these come from a single Israeli party. In addition, this party has no representatives in Parliament (despite having former members of Parliament on the candidate list). 
Unfortunately, we were unable to gain access to additional primaries data from Israel, with many of the parties declining our request due to privacy issues or the associated expenses of anonymizing the data. \new{Nevertheless, our unique dataset provides a rare opportunity to investigate the matter in question. In future work, we plan to seek additional primaries data, potentially from other countries, in order to further strengthen, and hopefully generalize, our results. Specifically, primaries data from parties of different ideological, social, and cultural societies may provide additional insights.} We also intend to investigate the attitudes towards adopting proportional ranking aggregation methods in other political parties by approaching party officials and proposing to implement the said methods for them to use. In another avenue, we plan to examine the use of other proportional representation methods intended at similar party tasks such as electing committees (a non-ranked subset of candidates) or electing representative for administrative roles (non-ranked yet ordered set of candidates).

% All data and code used by the Demokratit party are available at \url{https://github.com/elektronaj/democratit} in order to encourage other researchers and political parties to adopt and examine aggregation methods that promote proportional representation in  primaries. All other Python codes used in the course of this study will be provided upon request.

% All data and code used in this study are available at \url{https://github.com/BLIND_FOR_REVIEW} in order to encourage other researchers and political parties to adopt and examine aggregation methods that promote proportional representation in  primaries. 

% to tackle the important and challenging task of promoting proportional ranking.

% We hope that this work will encourage additional parties to adopt aggregation methods that promote proportional representation in their primaries. In order to allow The anonymized voter data and our Python code for executing the ? and ? examined aggregation methods are available at . \url{https://github.com/szufix/pabulib} 

% \cite{astudillo2021party} claimed that ‘quota’ system for party leadership would also solve supply-side problems regarding having women as party leaders since women’s presence is guaranteed. The issue of who would become the party’s top candidate remains, however, unresolved. 
% There is an overview here on the current status \cite{krook2010quotas}
% \cite{buckley2014will} shows that quaoas have positive representative effects in Ireland
% We can relate to that!

% \begin{acks}
% %
% Nimrod Talmon was supported by the Israel Science Foundation (ISF;
% Grant No. 630/19).
% %
% \end{acks}

\bibliographystyle{plain}
\bibliography{bib}
\end{document}